\documentclass[a4paper,11pt]{article}
\pdfoutput=1 

\usepackage{jinstpub} 
\usepackage{hyperref}
\newcommand{\MAPbBr}{L$_2[$MAPbBr$_3]$PbBr$_4$ } 
\newcommand{\FAPbBr}{L$_2[$FAPbBr$_3]$PbBr$_4$ } 
\newcommand{\MAPbI}{L$_2[$MAPbI$_3]$PbI$_4$ } 
\newcommand{\FAPbI}{L$_2[$FAPbI$_3]$PbI$_4$ } 
\newcommand{\CsPbI}{L$_2[$CsPbI$_3]$PbI$_4$ } 

\title{Light Yield of Perovskite Nanocrystal-Doped Liquid Scintillator}

\author[a,1]{E. Graham,\note{Corresponding author.}}
\author[a,b]{D. Gooding,}
\author[a]{J. Gruszko,}
\author[b]{C. Grant,}
\author[c]{B. Naranjo,}
\author[a]{and L. Winslow}

\affiliation[a]{Massachusetts Institute of Technology, Department of Physics and Laboratory for Nuclear Science,\\77 Massachusetts Ave Cambridge, MA 02139, USA}
\affiliation[b]{Boston University, Department of Physics, \\590 Commonwealth Avenue
Boston, MA 02215, USA}
\affiliation[c]{University of California Los Angeles, Department of Physics \& Astronomy,\\475 Portola Plaza, Los Angeles, CA 90095-1547, USA}

\emailAdd{lnor@mit.edu}

\abstract{Future generations of liquid scintillator neutrino experiments will require stably loading tons of candidate isotopes into kiloton-scale detectors while controlling the scintillator's optical properties. Nanoparticles' unique structural properties allow them to be used as highly-tunable wavelength shifters, which can be used to enhance double beta decay detection and background discrimination. Additionally, these nanoparticles can be made with double beta decay isotopes, which offers a promising method for isotope loading. Perovskite nanocrystals are particularly attractive due to the reliability of their crystal structure and their easily-scalable synthesis. We present here the first study of lead-based perovskite nanocrystals in a liquid scintillator experiment, demonstrating their properties as wavelength shifters and the scintillator cocktail's behavior under increasing nanocrystal mass-loading.}

\keywords{ Scintillators, scintillation and light emission processes (solid, gas and liquid scintillators); Particle identification methods; Large detector systems for particle and astroparticle physics; Neutrino detectors
}

\begin{document}
\maketitle
\flushbottom

\section{Introduction}
\label{sec:intro}

Liquid scintillator-based detectors have been at the center of many of the great discoveries of neutrino physics, from the first detection of neutrinos, to the discovery of neutrino oscillations, to current precision measurements with experiments such as KamLAND~\cite{kamReactor2013}, Borexino~\cite{borexinoSoloar2011}, Daya Bay~\cite{dayabay2016}, RENO~\cite{reno2018s}, Double Chooz~\cite{doublechooz2014} and PROSPECT~\cite{prospect2018}. The technology is well-suited to neutrino physics because it provides cost-effective scaling to kiloton-scale masses while providing straightforward background suppression through purification, spatial and temporal coincidence analyses, self-shielding, and pulse shape discrimination. While the achievements of current liquid scintillator experiments are impressive, the next generation of measurements has even more stringent background requirements and requires higher concentrations of dopants with a more varied set of isotopes. 

Liquid scintillator detectors are large calorimeters. The addition of directional information obtained by the timing-based separation of Cherenkov light from scintillation light would significantly enhance background suppression~\cite{direction2014, chessPRC, chessEPJC}. In kiloton-scale detectors, this separation can be improved by using a scintillator cocktail with a narrow emission spectrum centered at shorter wavelengths than traditional scintillator cocktails~\cite{direction2014}. Quantum dot-doped scintillators have inherently narrow emission spectra and are easily tuned through the geometry of the nanocrystals. They also provide a robust chemical means of suspending a variety of isotopes in the scintillator. 

Previous work has focused on scintillators doped with core-shell quantum dots~\cite{qdot1,qdot2}. In this paper, we investigate a promising new class of quantum dots, perovskite nanocrystals. These are also called perovskite nanoplatelets due to their flat geometry. These perovskite nanocrystals have fewer structural defects than core-shell quantum dots, producing more reliable spectra. Additionally, the synthesis of perovskite nanocrystals is far simpler, primarily using the technique of solvent precipitation. The simple and reliable synthesis of perovskite nanocrystals makes them excellent candidates for scaling to the amounts needed for a neutrino detector. Development of perovskite nanocrystals so far has produced nanocrystals with longer emission wavelengths that are less appropriate for detector applications. However, a detailed analysis of these long-wavelength perovskite nanocrystals will still motivate and inform future research and development for crystals that emit at shorter wavelengths.

This work builds on ref.~\cite{perovskite2016} and extends that work to explore properties relevant to detector applications. Following ref.~\cite{perovskite2016}, we verify the absorption and emission spectra of our scintillator cocktails to demonstrate successful synthesis. We then perform additional measurements relevant to the application of perovskite nanocrystals to particle detectors. We perform light yield measurements with a $^{22}$Na gamma source in order to determine the impact of perovskite nanocrystal doping on existing scintillator cocktails. We also investigate the concentration-dependent properties of the most promising perovskite nanocrystals, as identified during the light yield phase. Here, we extend beyond the modest concentrations employed in ref.~\cite{perovskite2016} to approach higher concentrations suitable for a neutrino physics experiment. We find the maximum loading of our perovskite nanocrystals and then determine the stability and brightness of this maximally loaded cocktail.

\section{Perovskite Nanocrystals}
\label{sec:perosvkites}

\begin{figure}[htbp]
\centering 
\includegraphics[trim=4.5cm 3.5cm 4.5cm 0cm, clip=true, width=0.45 \textwidth]{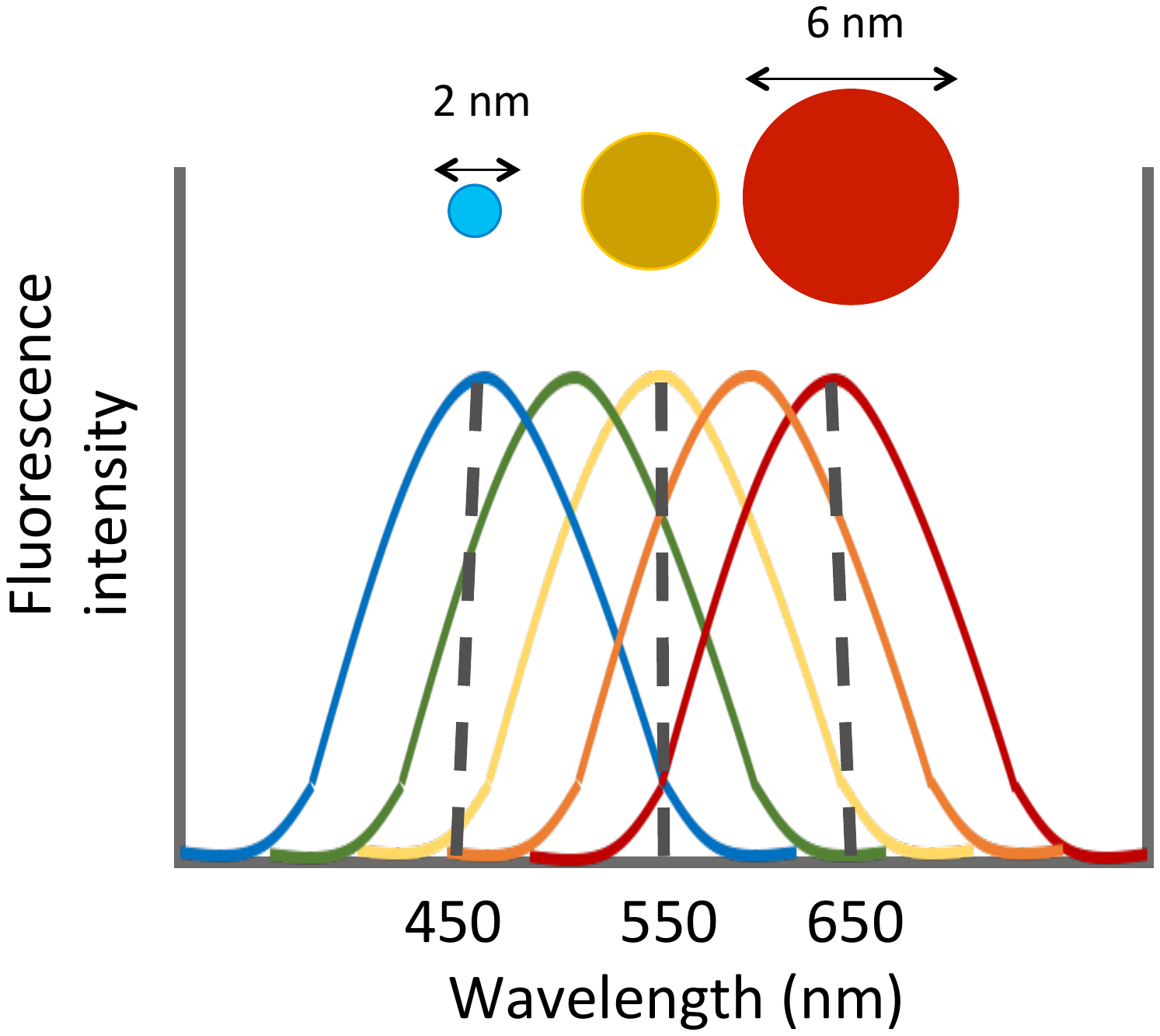}
\includegraphics[trim=2.25cm 2.5cm 4.5cm 3cm, clip=true, width=0.45 \textwidth]{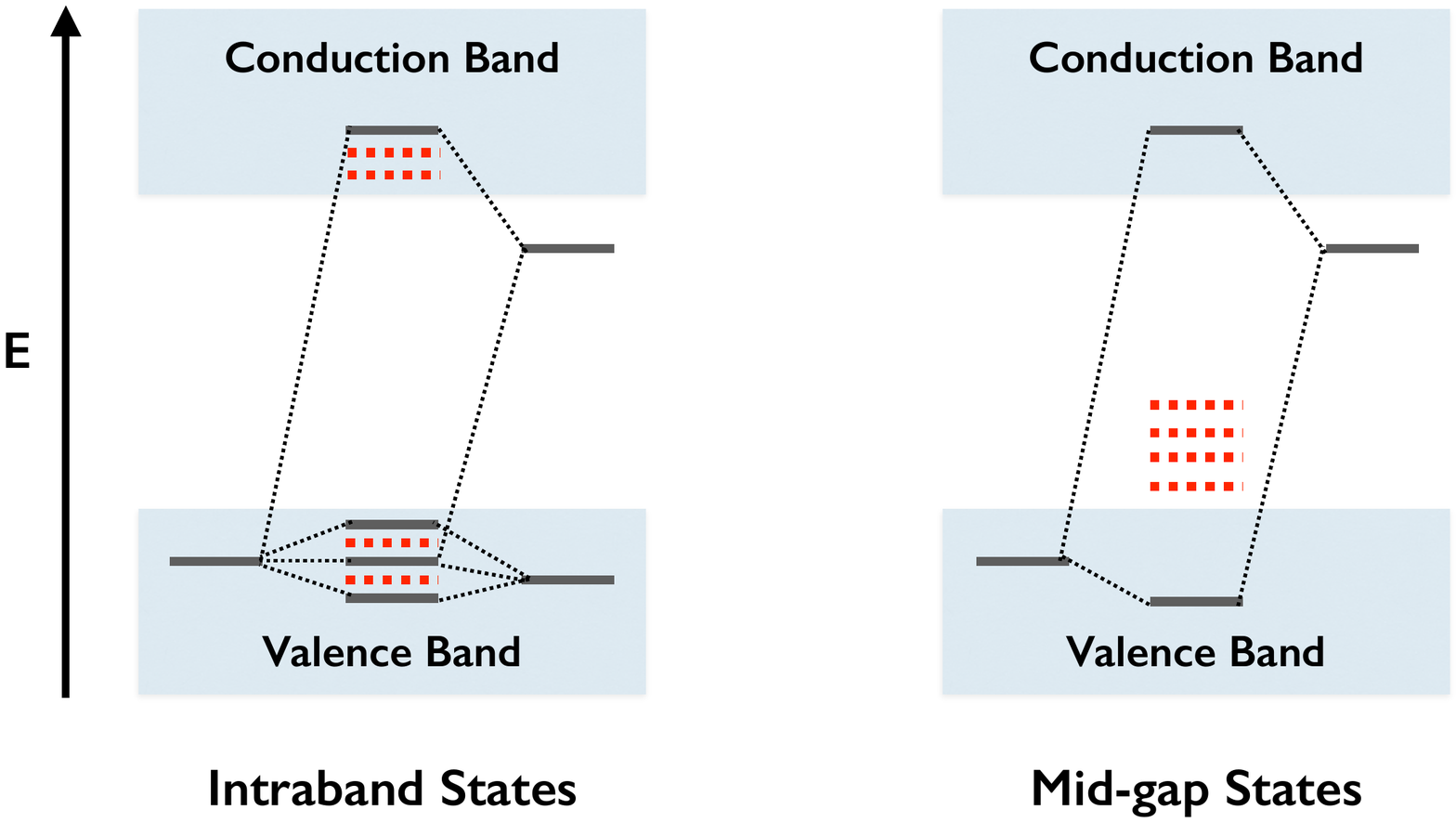}
\caption{\label{qdotIntro} \textit{Left:} Increasing the diameter of the nanocrystal increases the emission wavelength. \textit{Right:} Molecular orbital diagrams adapted from ref.~\cite{perovskites_Sun2018}. In the intraband case, common in canonical core quantum dots, defect states (red dashes) are contained in either the valence or conduction bands. Moreover, these states are shallow and transient. In the mid-gap case, common in perovskite nanocrystals, defect states form within the band gap. These states are not as shallow and decay non-radiatively, thus reducing the light output.}
\end{figure}

Quantum dots are semiconducting crystals with sizes on the order of nanometers. In this regime, the material's mobile charge carriers are effectively confined to a box. This quantum confinement causes the crystal's band gap to increase with decreasing crystal size, and thus blue-shifts its photoluminescence (see figure~\ref{qdotIntro}, \textit{left}). The tunable, size-dependent optical properties of quantum dots have applications in photocatalysis and solar cells, as well as in light emitting devices including televisions and LEDs \cite{nanocrystalEd}. 

The success of these applications requires highly luminescent crystals. Crystal defects such as dangling surface bonds and vacancies often appear in quantum dots. These defects create transition states within the bandgap (see figure~\ref{qdotIntro}, \textit{right}). This allows non-radiative recombination, thus reducing the light output. To combat these defects, the canonical core quantum dot can be encased in a protective shell made of another nanocrystal \cite{Reiss2009}. This encasing can be chosen to have the larger bandgap relative to the core crystal in order to shield the core from the formation of defects. Although the addition of a shell has been shown to increase the stability and luminescence of core quantum dots, these shelled dots are often difficult to fabricate, requiring high temperatures and inert atmospheres. Moreover, these procedures are difficult to control and the resulting core-shell quantum dots size often varies from batch to batch in both the extent of passivation and in the thickness of the shell.

\begin{figure}[t]
\centering 
\includegraphics[width=0.7\textwidth]{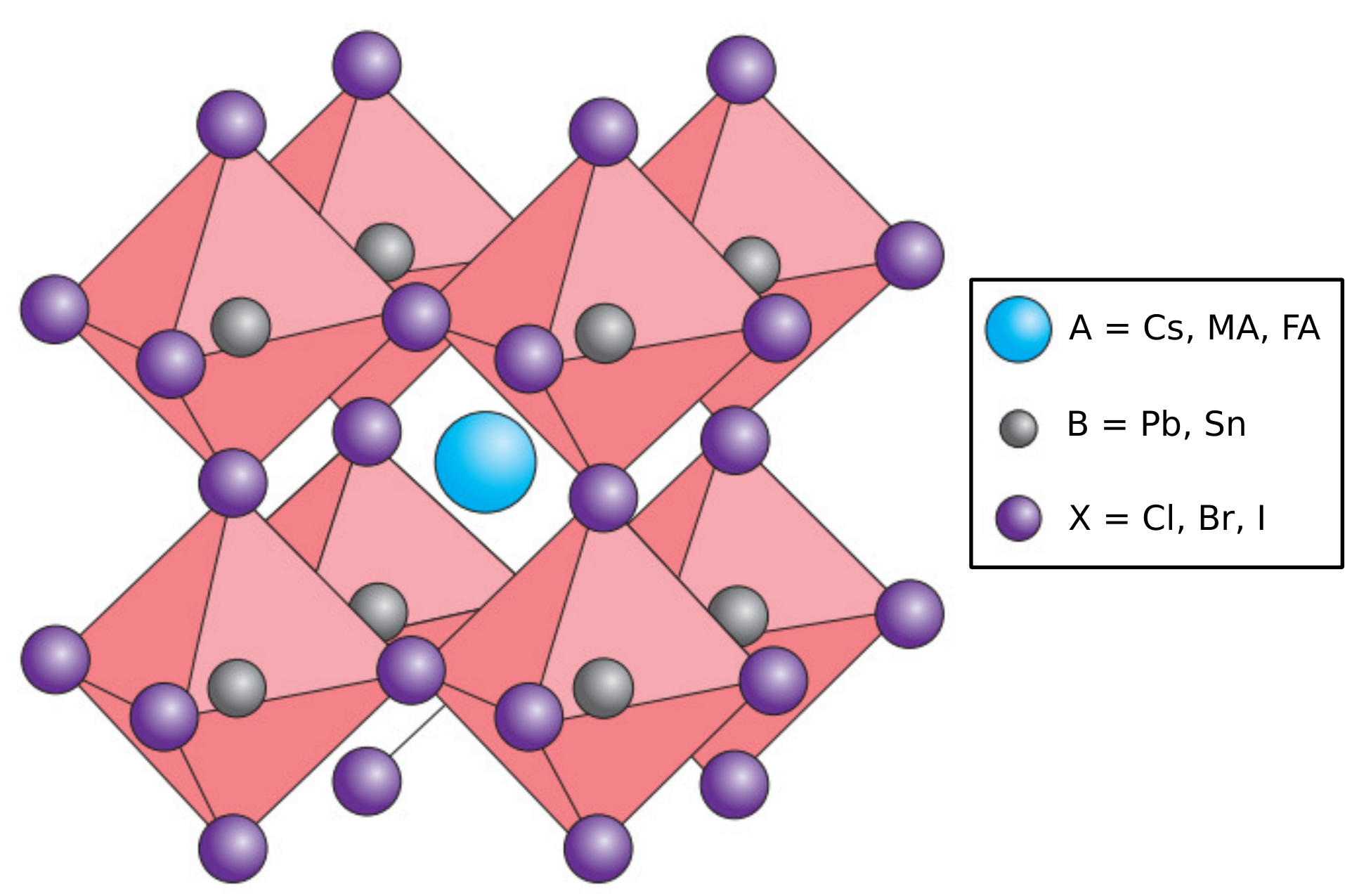}
\caption{\label{qdotJustCrystal} Perovskite nanocrystals have unit cells with structure ABX$_3$, where A is a monovalent organic or inorganic cation, B is a divalent metal cation, and X is a halide anion. These unit cells combine into double-layer perovskite nanoplatelets with ligands (not shown) blocking the formation of a bulk perovskite. In this work, we study lead\,(II) crystals with bromide and iodide halides and cesium, methylammonium, and formamidinium cations. Figure adapted from \cite{Green2014}.}
\end{figure}

A new type of quantum dot with a perovskite structure has emerged as a promising material, notable for its high luminescence and ease of synthesis \cite{Green2014}. The bulk perovskite crystal structure has the formula ABX$_3$, where A is a monovalent organic or inorganic cation, B is a divalent metal cation, and X is a halide anion (see figure~\ref{qdotJustCrystal}). The addition of ligands (L) in the synthesis prevents the formation of of the bulk perovskite, leaving nanoplatelets with thickness dependent on the starting ratio of LX:BX$_2$:AX. Changing the A, B, or X components of this crystal to smaller or larger atoms moves the emission wavelength to shorter or longer wavelengths. Moreover, these dots can be made by solvent precipitation, a more reproducible technique than hot-injection.

Perovskite nanocrystals are unique in their tolerance to defects on the crystal surface, even without passivation by a second nanomaterial. This property is due to the material's electronic properties and ionic bonding character, which cause defect states to form in either the valence or conduction band, rather than in the band gap \cite{Kim2014}. These intraband states come from the antibonding character of the valence maximum. This antibonding character causes nonbonding defect states to form below this maximum, and thus in the valence band. In the conduction band, the minimum is stable due to strong spin-orbit coupling, causing only shallow defect states to form. To first order, these non-bonding states are contained in the conduction band, leaving a clean bandgap. By contrast, defect-intolerant nanomaterials exhibit a valence band maximum with bonding character, causing defect states to form mid-bandgap. Both placements of defect states are shown in figure~\ref{qdotIntro} \textit{(right)}. Without strong spin-orbit coupling from the conduction band, these states tend to be less shallow, and thus less transient than intraband states. These mid-bandgap states undergo non-radiative recombination, decreasing the light output of the crystal.

The most-studied crystal structure is that of the lead-based perovskite nanocrystals, B\,$=$\,Pb\,(II), which is therefore the focus of this work \cite{Green2014}. The search for alternatives to lead that can be used in perovskites must consider the materials' electronic properties. The antibonding character of the valence maximum and strong spin-orbit coupling of the conduction band require two electrons in the outermost \textit{s} orbital. Antimony, bismuth, and tin fit this description, and their use is an active area of perovskite research \cite{Kamat2017}. Other more exotic atoms may also be possible and would be the subject of our group's future work.

\section{Perovskite Nanocrystal Synthesis} 

We synthesize perovskite nanocrystals using the method described in ref.\,\cite{perovskite2016}. Our solvents are purchased from Sigma Aldrich and our perovskite precursors from Great Solar Cell. We use them without further purification or drying. All syntheses are carried out in ambient lab conditions. 

\begin{figure}[htbp]
\centering 
\includegraphics[trim=0cm 4.5cm 0cm 4cm, clip=true, width=0.9\textwidth]{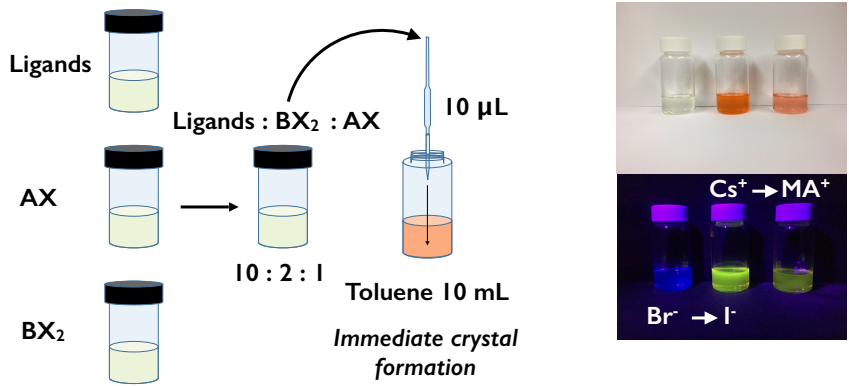}
\caption{\label{qdotSynthesis} \textit{Left:} A diagram of the synthesis, starting with precursors purchased from Great Solar Cell. \textit{Right:} Example samples (from left to right: L$_2[$MAPbBr$_3]$PbBr$_4$, L$_2[$CsPbI$_3]$PbI$_4$, L$_2[$MAPbI$_3]$PbI$_4$) in ambient light and under UV illumination.}
\end{figure}

The synthesis is summarized in figure~\ref{qdotSynthesis} \textit{(left)}. We first prepare a solution of the perovskite ligands by dissolving an n-butylammonium halide and n-octylammonium halide in dimethylformamide (DMF), each to a concentration of 0.5\,M. Two more solutions, one of halogenated A-site cation (AX)  and one of halogenated B-site cation (BX$_2$) are prepared, each at a concentration of 0.1 M in DMF.   The final precursor solution consisting of a 10:2:1 molar ratio of the ligands, BX$_2$, and AX  is mixed. After pipetting 10\,$\mu$L of this precursor solution into 10\,mL of toluene, perovskite nanoplatelets form immediately, as evidenced by photoluminescence upon radiation with UV light. This is shown in figure~\ref{qdotSynthesis} \textit{(right)}. The synthesis results in a scintillator cocktail with 3.8\,mg/L of Pb. This is consistent with the nanocrystal acting as a secondary wavelength shifter, rather than as a dopant. 

We synthesize perovskites from precursors containing two possible halides (bromide and iodide) and three possible A-site cations (cesium, methylammonium, and formamidinium). Of the six possible combinations, the only one not prepared is the brominated cesium perovskite; CsBr has poor solubility in DMF, and requires mixing with highly toxic dimethylsulfoxide (DMSO). 

\section{Fluorescence and Absorption Measurements}
\label{sec:spect_meas}
\begin{figure}[htbp]
\centering 
\includegraphics[width=0.7\textwidth]
{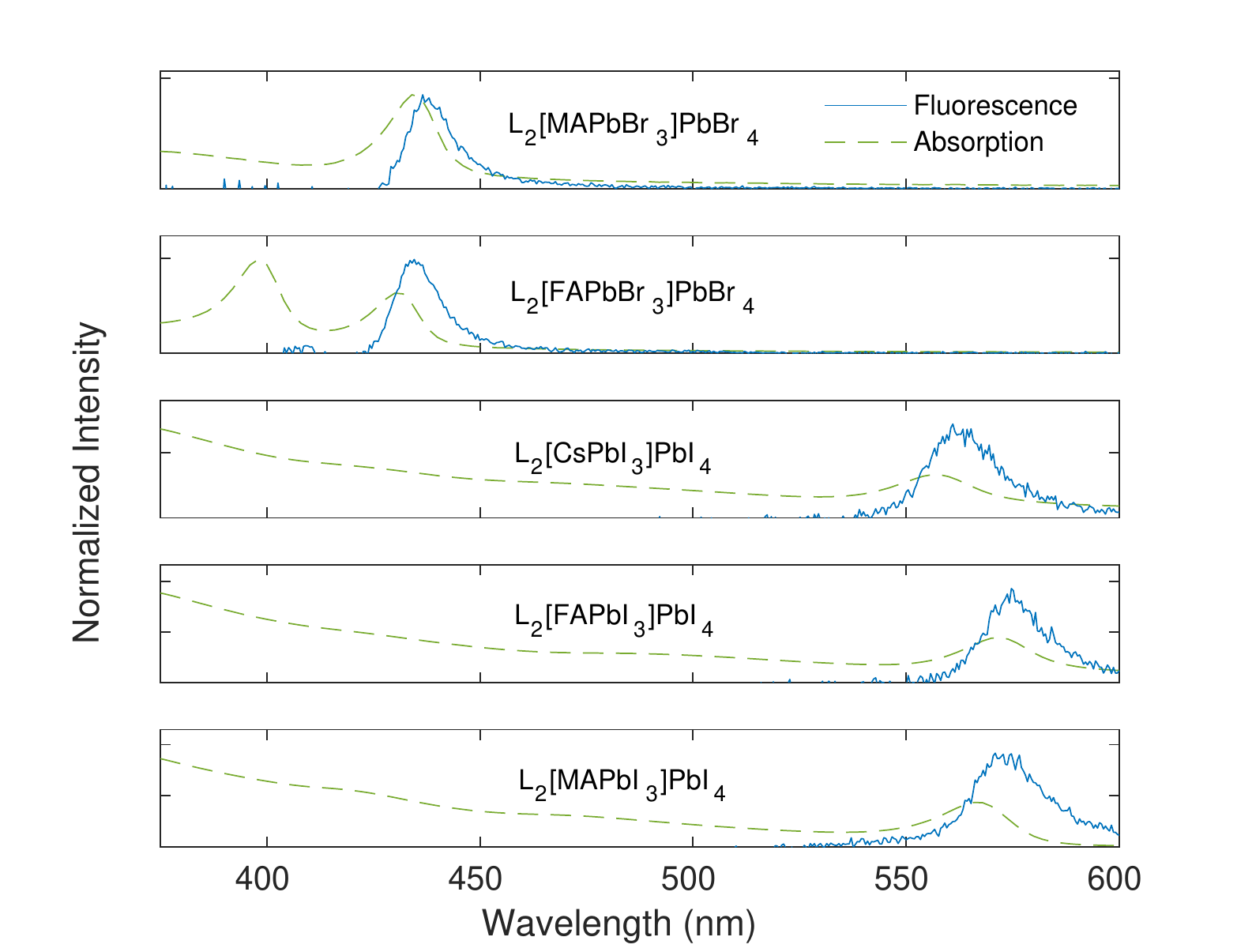}
\caption{\label{qdotEmissionAndAbsorption} The absorption and fluorescence emission spectra for the prepared samples. The fluorescence measurement features a narrower binning due to the resolution of the spectrophotometer.}
\end{figure}

We make fluorescence and absorption measurements in order to verify the success of our synthesis and to understand the optical properties of the perovskite nanocrystal and toluene cocktails. We add an additional scintillator to these cocktails to increase brightness and approximate actual detector cocktails: 2,5-diphenyloxazole (PPO) at a concentration of 1\,g/L. We use a Duetta spectrophotometer from HORIBA Scientific, which is capable of making both types of measurements. The results of these measurements are shown in figure~\ref{qdotEmissionAndAbsorption}.

The peak emission and absorption values are tabulated in table~\ref{tab1} and compared to those obtained in ref.~\cite{perovskite2016}. The peak emission and absorption values are well defined and are calculated by an interpolating Gaussian fit with an uncertainty of 0.1 nm. We add an additional linear term to the fit of absorption peaks to capture the underlying slope of the data.

\begin{table}[htbp]
\centering
\caption{\label{tab1} Results of fluorescence and absorption measurements. The quantum efficiency of a Hamamatsu model R13089 PMT at the fluorescence wavelength is provided for reference.}
\smallskip
\begin{tabular}{l c c c c c}
\hline
 & \multicolumn{2}{c}{Absorbance (nm)}&\multicolumn{2}{c}{Fluorescence (nm)} & PMT Quantum Eff.\\
Sample & Ref.~\cite{perovskite2016} & This work & Ref.~\cite{perovskite2016} & This work & Hamamatsu R13089\\
\hline
\MAPbBr & 431 & 434.6 & 437.3 & 438.1 & 0.23\\
\FAPbBr & 434 & 430.9 & 439 & 434.9 & 0.23\\
\CsPbI  & 553 & 557.1 & 561.1 & 562.7 & 0.06\\
\MAPbI & 566 & 567.7 & 573.9 & 573.2 & 0.04\\
\FAPbI & 566 & 572.0 & 575 & 574.8 & 0.04\\
\hline
\end{tabular}
\end{table}

The values for the primary emission and absorption peaks agree for all of the measured nanocrystals, with a slight shift towards shorter wavelengths in the spectrum of L$_2[$FAPbBr$_3]$PbBr$_4$. The iodide nanocrystals show high absorption at shorter wavelengths in addition to the well-defined peaks, extending into the emission region of the toluene/PPO blank used for absorption measurement. We also see a second absorption peak in the spectrum of \FAPbBr which we identify as the primary absorption of L$_2$PbBr$_4$ described in ref.~\cite{perovskite2016}. The presence of this peak indicates that single-layer L$_2$PbBr$_4$ platelets are present in the solution in addition to the double-layer \FAPbBr platelets. These thinner platelets could be either remnants from synthesis or represent breakdown of the \FAPbBr platelets. 

Based on the agreement of absorption and emission values, we claim successful synthesis of all five types of perovskite nanocrystals. Our synthesis of \FAPbBr also contains single-layer L$_2$PbBr$_4$ platelets, which we do not investigate separately.

\section{Light-yield Measurements}

\begin{figure}[htbp]
\centering  
\includegraphics[width=0.55\textwidth]{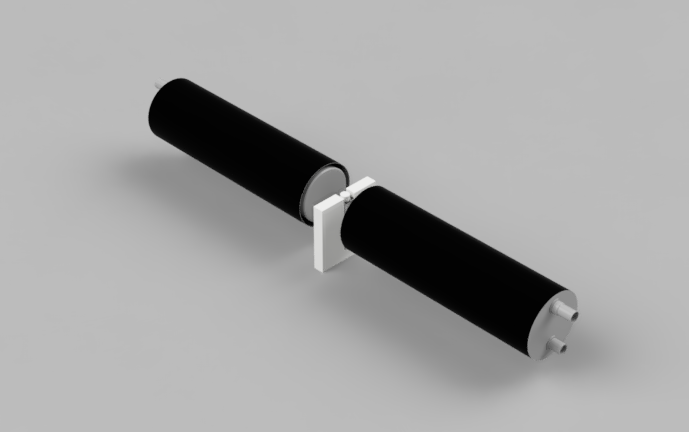}
\includegraphics[width=0.333\textwidth]{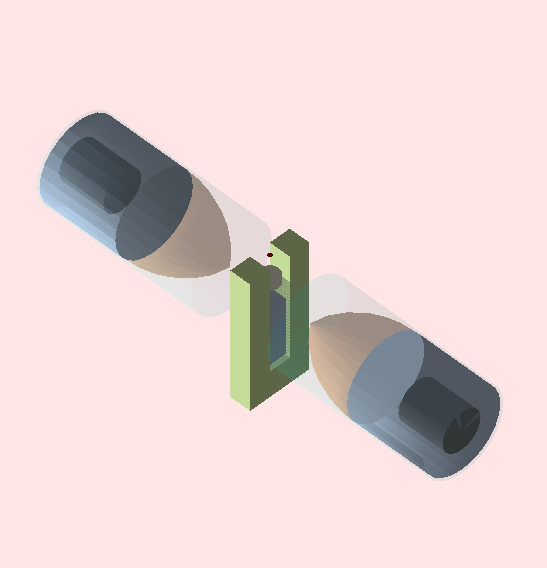}
\caption{\label{LightYieldApparatus} \textit{Left:} The light yield measurement stand, consisting of two PMTs, a quartz cuvette and a teflon cuvette holder. The cuvette holder and PMTs were held at a fixed distance of 1.8 cm apart using a system of clamps (not shown). \textit{Right:} The Geant4 geometry used for light yield simulation.}
\end{figure}

We begin our examination of phenomena relevant to scintillation detectors by measuring the light yield of our 10 $\mu$L cocktails for all samples. This serves as a verification of both our measurement techniques and the expected trends among the samples. The light yield of the scintillator cocktails is measured with an apparatus containing two photomultiplier tubes (PMTs) and a quartz cuvette, arranged as shown in figure \ref{LightYieldApparatus}. The PMTs are Hamamatsu model R13089, which is a high quantum efficiency 2-inch PMT. The PMTs are operated at 1900\,V, provided by a CAEN GECO DT5533EN high voltage module. The cuvette used is a Spectrosil 23-Q-10, with exterior dimensions of 12.5\,mm $\times$ 12.5\,mm $\times$ 48\,mm. 

The signal from each PMT is read out directly by the digitizer, a DRS4 evaluation board from PSI, sampling at 2\,GS/s. This digitizer is set to trigger on the coincidence of the two PMT signals, with a threshold of 2\,mV. The PMTs are roughly gain matched at the operating voltage. Each triggered signal is written to disk and converted to HDF-5 format on the data acquisition computer for offline processing. For each acquired waveform, the upper and lower bounds of the PMT pulse are identified using a pulse-finder based on the DEAP algorithm \cite{deap}. The total charge is then determined by integrating across the pulse-finder result, and converted to pC by dividing by the nominal 50\,$\Omega$ termination of the board. We perform our fits in terms of photoelectrons, using a gain calibration extracted from measurements of single-photoelectron dark pulses. In particular, we find that single-photoelectron pulses are 0.5 pC and 0.63 pC on the two PMTs.

A $^{22}$Na button source is placed on top of the cuvette and is used as the calibration source. $^{22}$Na is a $\beta^+$ emitter that decays to the first excited state of $^{22}$Ne, producing a 1.275\,MeV $\gamma$ from the de-excitation of $^{22}$Ne and two 0.511\,MeV $\gamma$'s from the annihilation of the $\beta^+$. Here, high-energy $\gamma$'s such as those from $^{22}$Na are preferable due to the higher numbers of photons that they generate. Performing fits on the Compton edges corresponding to the peaks, rather than the peaks themselves, allows us to compensate for the resulting loss of containment. 

In order to extract the light yield of the scintillator cocktail, a simulation is used to model the spectral shape. The cuvette, scintillator, source, teflon holder, and PMTs are modeled in a Geant4 simulation~\cite{geant4one,geant4two}. We interface with Geant4 through the RAT package, an analysis tool based on GLG4sim, and make use of RAT's models for scintillation and PMT physics \cite{RAT,glg4}. Absorption is modeled within the simulation, while Cherenkov effects are turned off. Both $\gamma$'s are simulated and the total number of photoelectrons collected in the PMTs is recorded.

The conversion from the simulated light yield apparatus to the real light yield apparatus is performed through a fit. This fit has three free parameters: a normalization, a linear calibration constant, and the width ($\sigma$) of the Gaussian energy resolution. The linear calibration constant converts between simulated photoelectrons of the benchmark sample and observed photoelectrons of any sample, with a resolution given by the Gaussian energy resolution rescaled by the same factor. The observed resolution goes as $1/\sqrt{\mathrm{(\#\:of\:photoelectrons})}$ relative to the light yield and is therefore allowed to float from fit to fit. At the benchmark light yield of the toulene/PPO blank, the resolution is given by 0.03 $\pm$ 0.01 measured PE per simulated PE.

\begin{figure}[htbp]
\centering  
\includegraphics[width=0.45\textwidth]{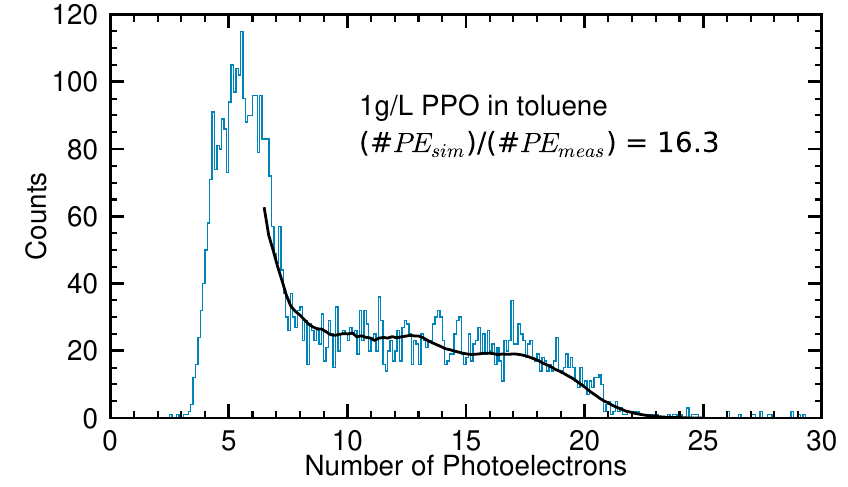}
\includegraphics[width=0.45\textwidth]{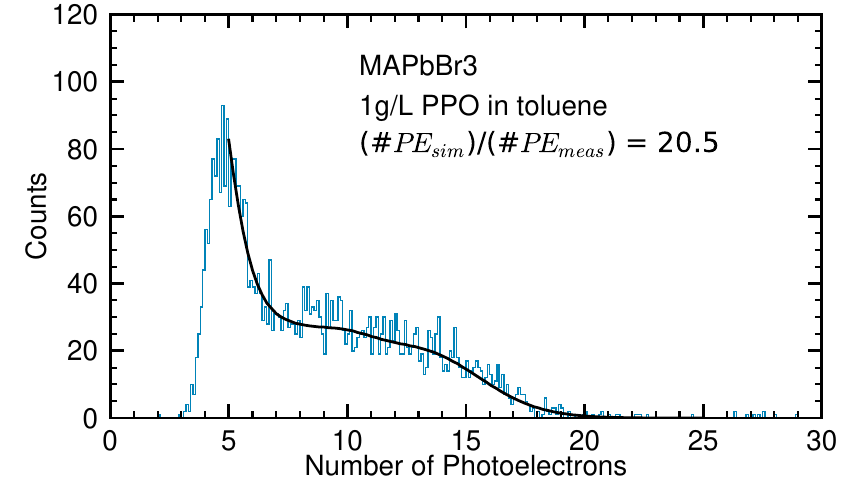}
\caption{\label{Spectrum} Sample spectra taken with the light yield apparatus: \textit{(left)} Toluene with 1\,g/L PPO and \textit{(right)} the \MAPbBr cocktail. The spectral shape is modeled using a Geant4 simulation of toluene with 1\,g/L PPO and fit to the spectrum with three free parameters to extract the light yield. Observed data is in blue while the fit is in black.}
\end{figure}

We benchmark against the calibrated ratio of simulated to observed photoelectrons for our toluene/PPO blank solution and note that reduced light yield of our perovskite solutions will manifest as a lower number of generated photoelectrons. We compare the number of observed photoelectrons for our perovskite solutions with the number of simulated photoelectrons for the benchmark solution, and extract a relative light yield by comparing the resulting calibration with the calibration of the benchmark solution. A subset of fits is shown in figure~\ref{Spectrum}.

The scintillating cocktails are made from toluene with 1\,g/L PPO and one of the five nanocrystal samples, just as in the fluorescence and absorption measurements. Light yield measurements are taken for each sample and for a benchmark cocktail, toluene with 1\,g/L PPO. The results of these fits are summarized in table~\ref{tab2}. The benchmark cocktail was measured several times to ensure reliable behavior of the apparatus. Its light yield was consistent to within fit uncertainties.

We also estimate the absolute light yield of the samples with the aid of our simulated apparatus, following the methods of ref.~\cite{LAB_ly}. First, we note that the light yield of the scintillator cocktail is set to 10000 photons/MeV in the simulation. We confirm this input light yield by dividing the number of generated scintillation photons in simulation by the corresponding visible energy, finding a value of 10100$\pm$800 photons/MeV. This indicates that no bias is introduced and that we can successfully recover the absolute light yield from our measurements. Then, we divide this simulated value by the linear constant described above, which converts between real photoelectrons and simulation photoelectrons, and extract our observed absolute light yield. These results are summarized in table~\ref{tab2} alongside the benchmarked results.

Our benchmark absolute light yield is on the same order of magnitude as the pure LAB sample described in ref.~\cite{LAB_ly}: 1000 photons/MeV. We stress that the primary goal of this measurement was not to make a case for the use of toluene with 1\,g/L PPO as a scintillator candidate, but rather to examine the impact of using quantum dots as secondary wavelength shifters in an existing scintillator. Many factors in our experimental setup, such as the presence of oxygen in our samples and the age of our PPO, may be negatively impacting the measured benchmark light yield. In particular, the presence of oxygen in PPO-based organic scintillators has been shown to decrease light yield by a factor of 20\% \cite{oxygen_LY}. However, the factors contributing to the light yield degradation are consistent between our benchmark solution and our preovskite nanocrystal solutions.

Using our Geant4 model of the measurement apparatus, we also make rudimentary simulations of the light yield of each perovskite nanocrystal. We set the emission spectrum of our scintillator cocktail in simulation to the emission spectrum of each perovskite nanocrystal, measured as described in section~\ref{sec:spect_meas}, and run our simulation otherwise unchanged. Then, we perform a linear fit of the resulting photoelectron distribution to the photoelectron distribution of the toluene/PPO simulation. As seen in table~\ref{tab2}, this simulation reproduces the general trends seen in the measurement apparatus but overestimates the light yield of the bromide nanocrystals. Since this simulation is only adjusted via the changed emission spectrum and not other optical properties such as absorption or scattering, some deviation is expected.

\begin{table}[htbp]
\centering
\caption{\label{tab2} Results of light yield fits for five types of perovskite nanocrystals in simulation and experiment, presented both as relative light yields benchmarked against toluene with 1\,g/L PPO and as absolute light yields. Fit uncertainties are provided for each.}
\smallskip
\begin{tabular}{c c c c c c}
\hline
Sample & Simulated LY & Measured LY  & Absolute LY [photons/MeV] \\
\hline
Toluene with 1\,g/L PPO & 1.00 & 1.00 & 620 $\pm$ 50 \\
\MAPbBr & 0.97 $\pm$ 0.01 & 0.795 $\pm$ 0.009 & 490 $\pm$ 40 \\
\FAPbBr & 0.88 $\pm$ 0.01 & 0.755 $\pm$ 0.010 & 470 $\pm$ 40 \\
\CsPbI  & 0.37 $\pm$ 0.01 & 0.353 $\pm$ 0.010 & 220 $\pm$ 20  \\
\MAPbI  & 0.32 $\pm$ 0.005 & 0.431 $\pm$ 0.007 & 270 $\pm$ 20 \\
\FAPbI  & 0.35 $\pm$ 0.005 & 0.448 $\pm$ 0.009 & 280 $\pm$ 20 \\
\hline
\end{tabular}
\end{table}

\begin{figure}[htbp]
\centering 
\includegraphics[width=0.7\textwidth]
{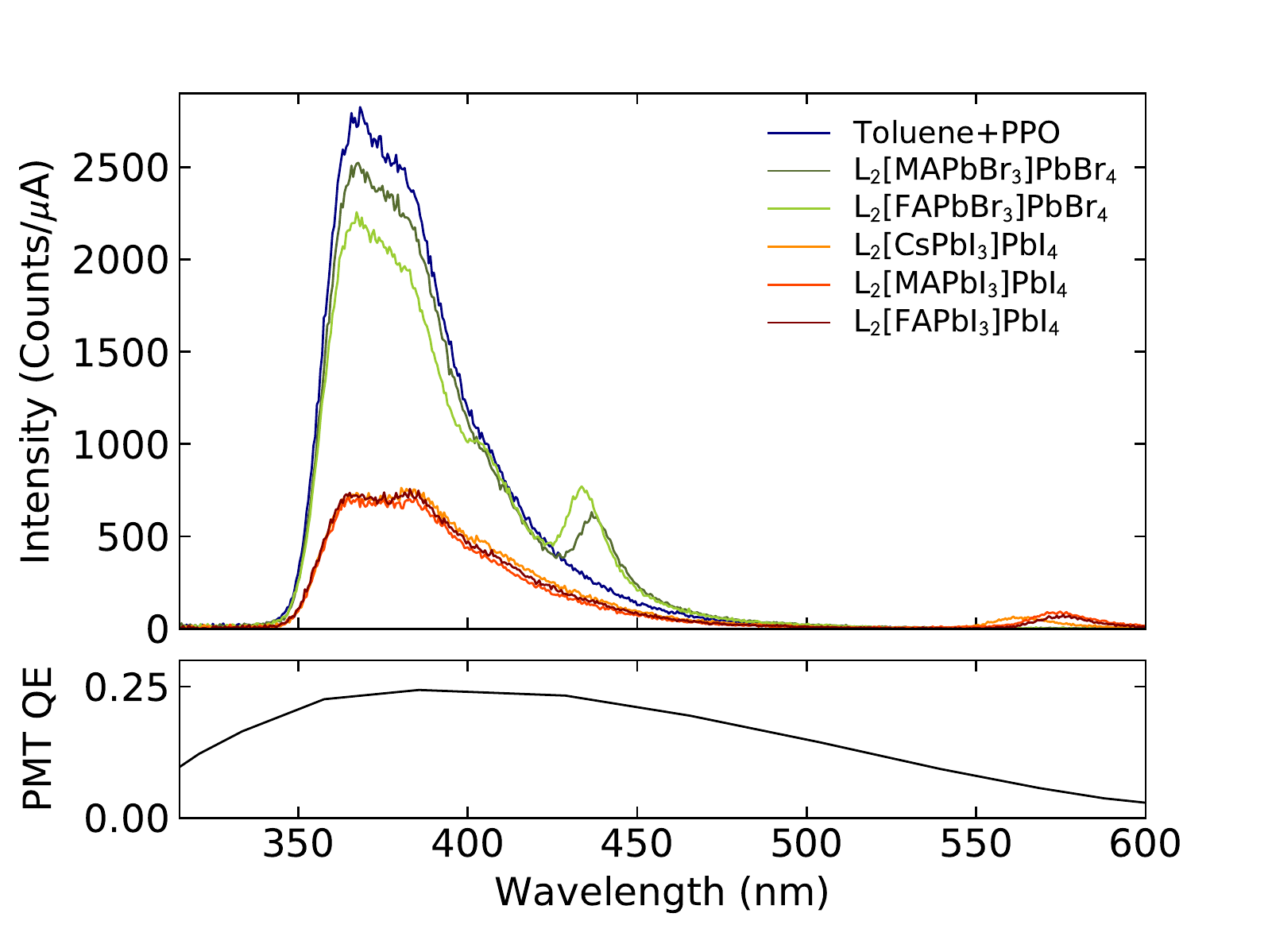}
\caption{\label{all_em} The emission peak of toluene/PPO for the blank solution and solutions containing each perovskite nanocrystal, showing a pronounced decrease when iodide nanocrystals are present. The emission peak of each nanocrystal is also present but small compared to the toluene/PPO emission. The quantum efficiency curve of a Hamamatsu R13089 PMT is provided for reference.}
\end{figure}

The addition of perovskite nanocrystals to the scintillator cocktails decreases the light yield. This is to be expected due to the presence of the heavy metal Pb in the perovskite nanocrystals, which increases the absorption of the cocktail at shorter wavelengths. This effect is the most pronounced for the iodide nanocrystals which emit around 570 nm, but is also present in the bromide nanocrystals. At the low concentrations used in this set of measurements, the light yield behavior is dominated by the magnitude of scintillation at the original 380 nm toluene/PPO peak. As demonstrated in figure~\ref{all_em}, the magnitude of toluene/PPO emission remains high for the bromide nanocrystals but drops for the iodide nanocrystals. This is consistent with the high absorption previously seen for the iodide nanocrystals at short wavelengths.

It is clear that much higher concentrations of perovskite nanocrystals are needed to shift the primary emission of the scintillation cocktail. The low light yield of the iodide nanocrystals can potentially be interpreted as a more effective shift of primary emission, because more of the primary toluene/PPO peak is masked. However, we do not pursue a high concentration study of iodide nanocrystals here; their long emission wavelengths are outside of the peak efficiency range of our PMT detectors. 

\section{High-concentration Loading of \MAPbBr}

The measurements detailed in the previous section examine solutions in which 10 $\mu$L of precursor solution is introduced into 10 mL of toluene/PPO cocktail, but this concentration is small relative to the requirements of most applications, including neutrinoless double-beta decay searches. Additionally, low concentrations of perovskite nanocrystals fail to fully absorb the primary emission of the toluene/PPO cocktail at short wavelengths, impairing the nanocrystals' function as wavelength shifters. In other words, their efficiency is low at this reduced concentration.

Using the spectrophotometer, we take fluorescence emission spectra of \MAPbBr in amounts ranging from 10 $\mu$L to 1 mL, with the toluene/PPO cocktail volume fixed at 10 mL, i.e concentrations ranging from 3.8 to 380 mg/L of Pb. The maximum amount of \MAPbBr that can be put in solution under these conditions is 1.3 mL (490 mg/L of Pb), at which point the solution rapidly turns clear as the perovskite nanocrystals break apart and cease emission. 

\begin{figure}[htbp]
\centering  
\includegraphics[width=0.8\textwidth]{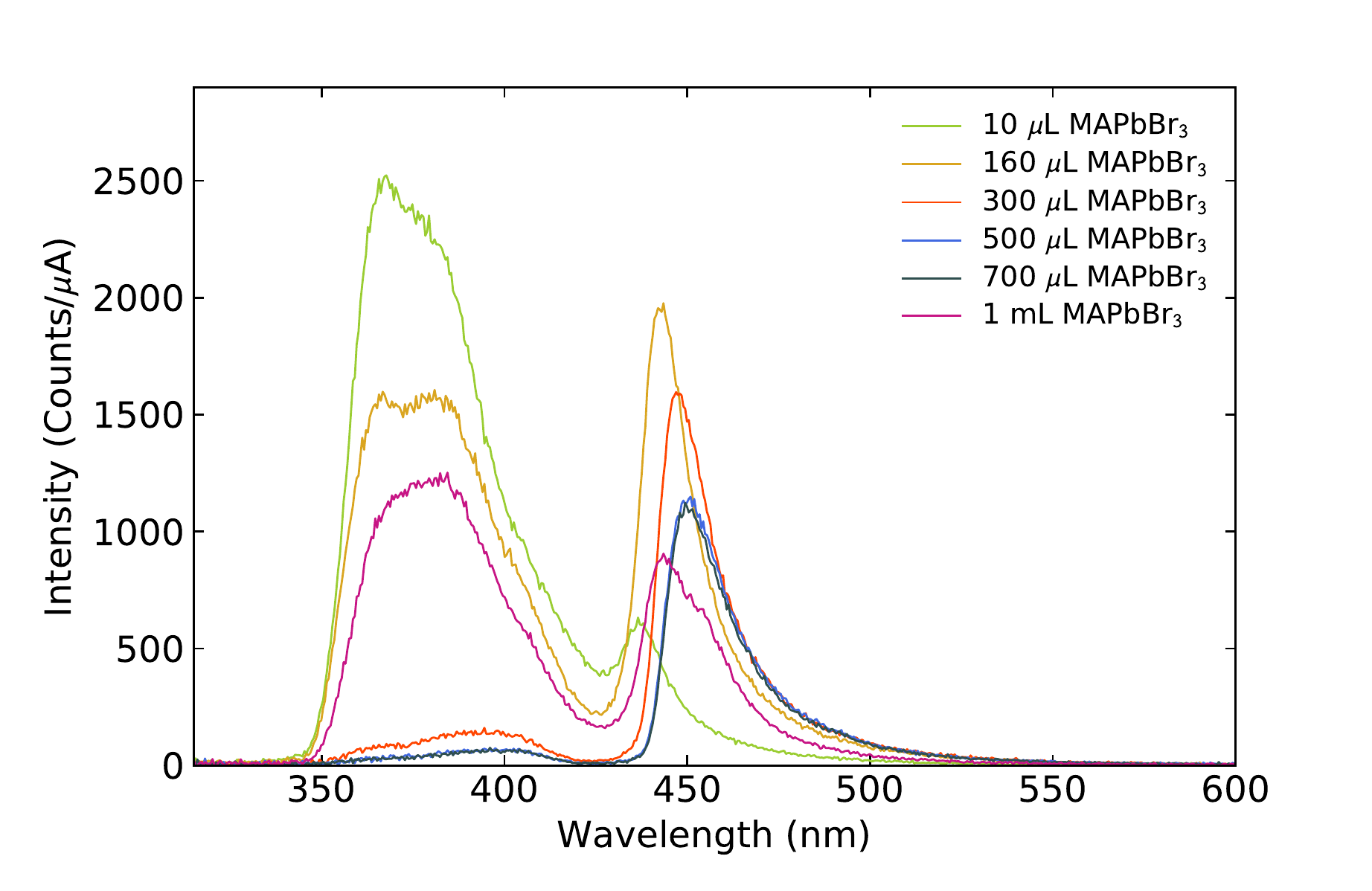}
\caption{\label{conc} Emission spectra for varied concentrations of \MAPbBr in 10 mL of toluene/PPO.}
\end{figure}

The emission spectra for a selection of \MAPbBr concentrations are shown in figure~\ref{conc}. The magnitude of the broad 380 nm toluene/PPO peak decreases for increasing loads up to 500 $\mu$L (190 mg/L of Pb), until the spectrum is dominated by the \MAPbBr response. However, the toluene/PPO peak reemerges for loads above 700 $\mu$L (270 mg/L of Pb), indicating that the solubility of \MAPbBr suffers at higher concentrations.

We observe a wavelength redshift, again visible in figure~\ref{conc}, with increased concentration of L$_2[$MAPbBr$_3]$PbBr$_4$. This effect was also seen in ref.~\cite{perovskite2016}, where it was linked to the number of distinct drops of precursor ligands added to the toluene solution. The 1 mL (380 mg/L of Pb) sample does not show this effect immediately after creation, instead resting at a redshift roughly equivalent to the 160 $\mu$L (61 mg/L) sample, but features a much broader peak that develops a greater redshift over time. The observed redshift has implications for large-scale synthesis of perovskite nanocrystals, suggesting the need for controlled deposition of drops of precursor solution.

\begin{figure}[htbp]
\centering  
\includegraphics[width=0.8\textwidth]{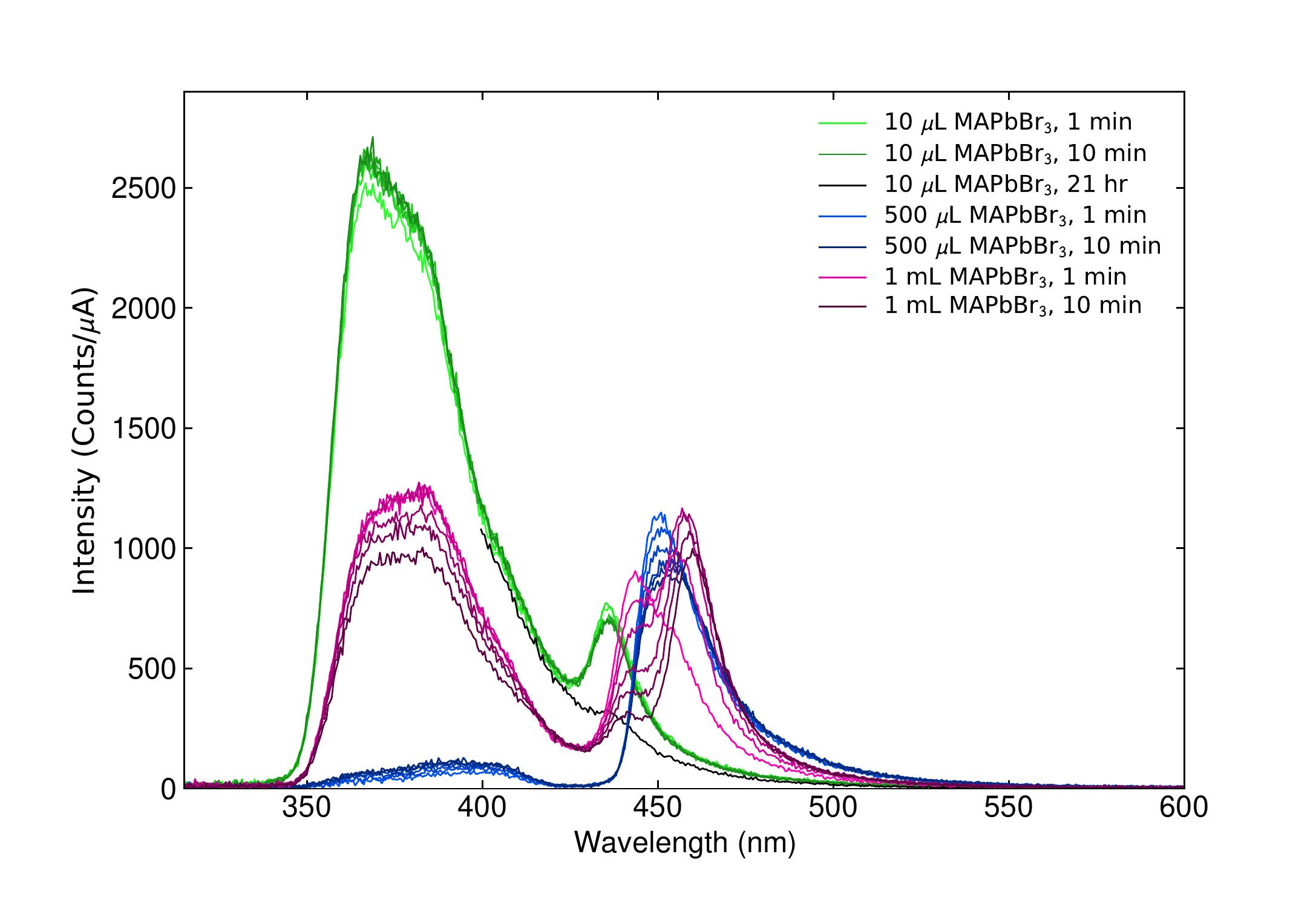}
\caption{\label{time} Emission spectra for varied concentrations of \MAPbBr in 10 mL of toluene/PPO over a period of ten minutes. Six spectra are plotted for each sample at intervals of two minutes. An additional spectrum for 10 $\mu$L of \MAPbBr after 21 hours is also shown.}
\end{figure}

The time-dependent behavior of the 1 mL (380 mg/L of Pb) solution suggests possible instability of our higher-concentration samples, so we perform a temporal characterization of our most promising samples. The behavior of 1 mL (380 mg/L of Pb), 500 $\mu$L (190 mg/L of Pb), and 10 $\mu$L (3.8 mg/L of Pb) solutions over a period of ten minutes is displayed in figure~\ref{time}. The time-dependent redshift of 1mL (380 mg/L of Pb) solution is clearly visible, but the 10 $\mu$L (3.8 mg/L of Pb) solution is highly stable over a period of ten minutes. 500 $\mu$L (190 mg/L of Pb) displays a small loss in peak magnitude and features time-dependent peak broadening, but is much more stable than the 1 mL (380 mg/L of Pb) solution. Finally, as shown by the spectrum plotted in black, the 10 $\mu$L (3.8 mg/L of Pb) sample is not stable overnight under ambient conditions. This is expected, as the sample was not isolated from light or oxygen, as would need to be done in a full test of stability.

A full determination of optimal loading also requires understanding the effects of high loading on the light yield of scintillator cocktails. In practice, we found that higher-concentration solutions, particularly the studied 500 $\mu$L (190 mg/L of Pb) solution, produced light below the single-photoelectron level in the geometry used in these measurements. Because of this, we instead estimate the light yield of high-concentration solutions using the simulation method discussed in the previous section.

\begin{figure}[htbp]
\centering  
\includegraphics[width=0.8\textwidth]{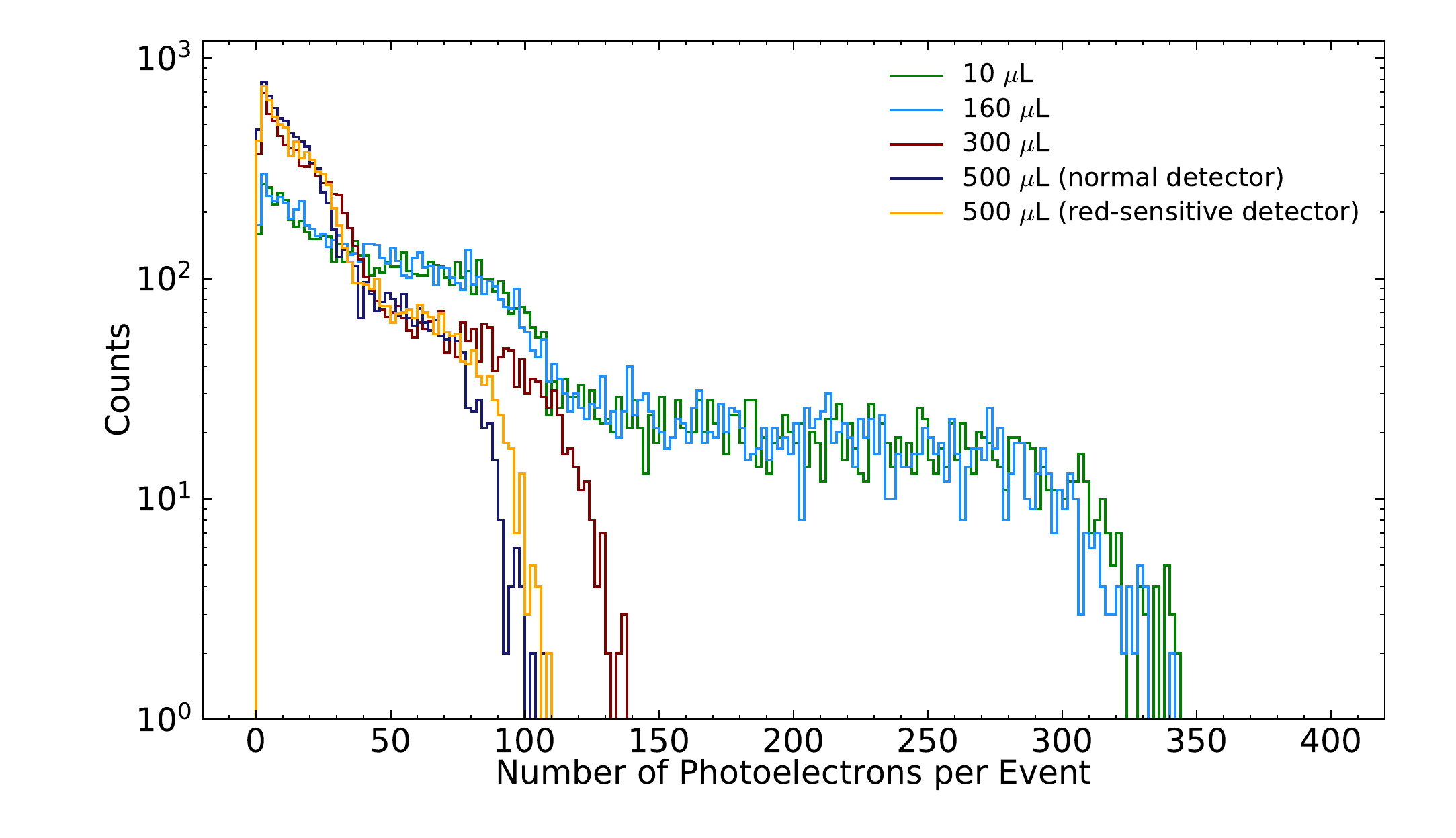}
\caption{\label{conc_ly} Geant4 simulation of photoelectron yield for varied concentrations of \MAPbBr in 10 mL of toluene/PPO, using experimentally measured emission spectra.}
\end{figure}

High-concentration solutions of \MAPbBr have lowered light yields, but intermediate concentrations do not show this effect. Photoelectron distributions from simulations performed using the experimental emission spectra of each concentration are shown in figure~\ref{conc_ly}. The 10 $\mu$L (3.8 mg/L of Pb) and 160 $\mu$L (61 mg/L of Pb) solutions have nearly identical photoelectron spectra, while the light yield drops for higher concentrations. The absolute light yield of the low-concentration samples found in experiment, as discussed in section 5, is 490 $\pm$ 40 photons/MeV. 

For the 500 $\mu$L (190 mg/L of Pb) spectrum, we perform an additional simulation using a hypothetical red-sensitive PMT with a constant quantum efficiency of 0.24 between 400 and 600 nm. This allows us to correct for loss of photoelectrons due to the reduced PMT efficiency in the \MAPbBr emission region. As expected, this correction produces a small increase in the amount of photoelectrons successfully registered by the PMT in simulation. However, even with this correction, the light yield is still lower than that seen at low concentrations. Future measurements will evaluate whether this reduced light yield is viable for particle detection and investigate chemical and environmental means of potentially increasing this light yield.

\section{Conclusion}
In this work, we successfully synthesize five types of perovskite nanocrystals. We measure their light yield benchmarked against 1 g/L toluene/PPO solution, both in simulation and in experiment. At the dilute concentrations used in our original synthesis, the light yield is dominated by the toluene/PPO emission of the scintillator cocktail, rather than the perovskite nanocrystals. However, the high absorption of iodide nanocrystals reduces the toluene/PPO emission by a factor of three. The bromide nanocrystals show more modest reductions in light yield and are also more promising due to the short wavelength of their emission peaks.

We also study the concentration-dependent behavior of L$_2[$MAPbBr$_3]$PbBr$_4$, one of the bromide nanocrystals. We determine that in ambient lab conditions the maximum stable loading is 500 $\mu$L in 10 mL of toluene (190 mg/L of Pb), while the absolute maximum loading is 1.3 mL (490 mg/L of Pb). Perovskite nanocrystals are far more stable in light-tight, oxygen-free, and low-humidity environments, leaving open the possibility that greater loadings and higher stabilities could be achievable under operating conditions that are already common in low-background experiments. This remains a topic for future work. Finally, we consider the light yield of various concentrations of \MAPbBr in simulation. We find that the light yield remains stable up through 160 $\mu$L (61 mg/L of Pb) but drops for higher concentrations, even when corrections are performed for PMT efficiency. 

Advancements in the chemical synthesis of perovskite nanocrystals will be necessary for future applications in particle detectors. The nanocrystals studied in this work are lead-based, and therefore have high intrinsic background levels. We focused on these materials because they are the oldest and most well-studied of the perovskites; however, extensive work is being done within the materials science research community to develop lead-free perovskites \cite{Pbfree_perovskites}. Two promising candidates are tin and germanium, both of which have double beta decay isotopes. These candidate nanocrystals could allow for simpler and higher-mass isotope loading in future liquid scintillator-based double beta decay experiments, without increasing intrinsic background levels. 

 Similarly, higher light yields and improved stability will be necessary for successful detector operation. Some improvement in both metrics is expected from deoxygenating the solution and keeping all materials in a nitrogen-purged environment, two practices that are commonly used for liquid scintillator experiments, and will be used in our future studies. Stability and improved efficiency in more challenging settings, like those found in devices, is a major focus of the perovskite research community \cite{stability_perovskites}, and we expect many of the improvements developed for those purposes to be directly applicable to uses in fundamental physics. 

Perovskites could also provide a benefit to future low-background liquid scintillator experiments by serving as highly-tunable wavelength shifters, as shown by the variety of spectra seen in the nanocrystals studied here. Shifting the emission and absorption spectra of scintillators to shorter wavelengths would help efforts to separate Cherenkov and scintillation light in next-generation detectors \cite{direction2014}. Since perovskites are under active study for use in devices like solar cells and light-emitting diodes, major research efforts are focused on improving making high-efficiency and stable blue- and ultraviolet-emitting perovskites \cite{blue_perovskites}. 

Though as a technology, perovskites can still be considered to be in their infancy, we have demonstrated for the first time that they can be used a wavelength shifters in a liquid scintillator experiment. Many improvements are needed before they can be considered a viable candidate for loading in large-scale experiments, but we have shown that their synthesis and use is simple enough to be performed with minimal equipment, and could be conducted in an underground laboratory setting if necessary. Significant resources are currently being devoted to improving perovskite performance and scaling up their synthesis to industrial scales, developments that could be used to improve future kiloton-scale liquid scintillator-based double beta decay experiments.

\acknowledgments
This work is supported by NSF award number 1554875 and 1806440. We thank Watcharaphol Paritmongkol for advice on the synthesis. The authors thank Riccardo Comin, Jake Siegel and William Tisdale for helpful discussions on perovskite nanocrystals. This work is part of the NuDot experiment and we thank that collaboration for many useful discussions.

\bibliographystyle{JHEP}
\bibliography{main_perovskite.bib}

\end{document}